\documentclass[twocolumn,pre,showpacs]{revtex4}
\usepackage{graphicx}
\usepackage{epsfig}
\usepackage{bm}

\begin{document}

\title{Two-Dimensional Diffusion in the Presence of Topological Disorder}

\author{Ligang Chen}
\author{Michael W. Deem}

\affiliation{
Department of Physics \& Astronomy
\\
Rice University, Houston, TX  77005--1892}

\begin{abstract}
How topological defects affect the dynamics of particles
hopping between lattice sites of a distorted, two-dimensional
crystal is addressed.  Perturbation theory and numerical simulations
show that weak, short-ranged topological 
disorder leads to a finite reduction of the
diffusion coefficient.  Renormalization group theory and numerical
simulations suggest that longer-ranged disorder, such as that from
randomly placed dislocations or random disclinations with no net
disclinicity, leads to subdiffusion at long times.
\end{abstract}

\maketitle

\section{Introduction}

Diffusion in random media is a well-studied problem
\cite{Bouchaud}.
The mean-square displacement of a tracer particle
behaves at long times in a way that depends on the character
of the random forces induced on the tracer by the
disorder.  Forces that arise from random potentials
lead to a reduction of the transport, with subdiffusion possible
for diffusion of an ion in a medium with quenched charges
obeying bulk charge neutrality \cite{Bouchaud}.
Interestingly, the same subdiffusion results from
diffusion of an ion in a medium with randomly-placed, quenched
dipoles \cite{Nelson1,Nelson2,Cha,Deem3}.
Forces that arise from entrainment along
fluid streamlines lead to an increase
in the transport, with the well-known result of 
turbulent super-diffusion
possible for random streamlines with statistics
characteristic of fluid turbulence \cite{Bouchaud}.

Distortion of the underlying lattice upon which the diffusion
occurs is a very different type of disorder.
In particular, topological defects such as
dislocations or disclinations should affect the
transport properties of a diffusing tracer particle.
These topological defects cause a global rearrangement of
the connectivity of the lattice upon which the diffusion occurs.
Moreover, there is an elastic response of the
lattice to such defects, and so there is also local
expansion or compression of the crystal unit cells.
Study of how such topological defects affect the
transport is, therefore, an interesting 
and challenging problem.
Among other results,
it might be expected that randomly-placed dislocations and
random disclinations with no bulk disclinicity will
lead to similar dynamics,
given the results regarding dynamics in random potentials and
the analogy between linear elasticity theory and
electrostatics.

Previous work has begun to address the question of how
topological disorder affects the transport.
Random disclinations, with no net disclinicity, were
predicted to lead to subdiffusion \cite{Turski}.
A single dislocation, on the other hand, was predicted to
increase the local diffusivity \cite{Turski2}.
These studies, however, were approximate \cite{Kleinert}.
In particular, rotational symmetry was assumed in the
dislocation problem, and no effects of lattice
expansion or contraction were allowed in the
disclination problem.

Transport in a two-dimensional crystal with topological defects, then,
remains an interesting and unsolved problem.
Our model of surface diffusion, and the Fokker-Planck
equation that results, is introduced in Sec.\ II.
How the topological defects affect the transport,
and a field theoretic description used to analyze the dynamics,
is described in Sec.\ III.
Perturbation theory and computer simulation are used to examine
the effect of non-singular topological disorder on the diffusion coefficient
in Sec.\ IV.  The possibility of anomalous diffusion in singular
topological disorder is examined by renormalization group theory and
computer simulation in Sec.\ V.  A discussion of the results, and their
relation to the previous literature, is given in Sec.\ VI.
A discussion of the effects of torsion, which exists solely within
the cores of defects, is given in Sec.\ VII.
We conclude in Sec.\ VIII.

\section{The Surface Diffusion Model}

We consider a particle hopping on the surface of a crystal. 
The particle hops only between nearest-neighbor lattice sites,
and the rate of hopping is constant.  In particular, since
surface diffusion is usually an activated process,
the rate to
hop between neighboring sites is assumed to be independent of the
distance between sites.
Disorder in the spatial arrangement of the surface
lattice sites indirectly affects the 
diffusion dynamics through modification of the hopping events.

We derive the Fokker-Planck, or diffusion, equation for the
surface species by two independent methods.  In the first
method, the hopping dynamics is derived from
a physically-motivated consideration of the master equation
for the process.  In the second method,
the result is derived in an efficient
fashion by considering a change of variables in the field theory for the
dynamics.

The particle is considered to hop
on an irregular grid of lattice sites.  The probability for particles
to be on a given site, $P({\bf r})$, decreases with time due to hopping of
particles off the site  and increases
with time due to hopping of neighboring  particles onto the 
site (see Figure \ref{fig1}):
\begin{figure}[tbp]
\centering
\leavevmode
\psfig{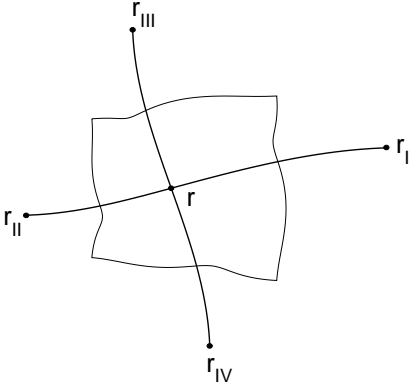}
\caption{A lattice site, ${\bf r}$,
on the distorted crystal and the four nearest
neighbors are shown schematically.  Also shown is the 
distorted unit cell of the
central lattice site.}
\label{fig1}
\end{figure}
\begin{eqnarray}
{d [V({\bf r}) P({\bf r}, t})]&=& \frac{D_0 \Delta t}{h^2} \bigg[
V({\bf r}_I) P({\bf r}_{I}) + 
V({\bf r}_{II}) P({\bf r}_{II}) + 
\nonumber \\ && + 
V({\bf r}_{III}) P({\bf r}_{III}) + 
V({\bf r}_{IV}) P({\bf r}_{IV}) 
\nonumber \\ &&
- 4 V({\bf r}) P({\bf r}, t)
\bigg] \ ,
\label{1}
\end{eqnarray}
where $h$ is the lattice spacing, $\Delta t$ is the small time increment, and
$D_0$ is the diffusion coefficient.  The volume of each, possibly distorted,
 unit cell is given by $V({\bf r})$.
Equation (\ref{1}) is exact and leads in the continuum limit
to the general expression for diffusion in curved space \cite{Ikeda}.
Although the crystal may be distorted, a regular crystal lattice
can always be defined locally 
in terms of lattice coordinates $\mbox{\boldmath $\sigma$}$.
In the $\mbox{\boldmath $\sigma$}$ space, the particle hops
either up, down, left, or right.
  The correspondence
is given by 
${\bf r}_I \leftrightarrow {\mbox{\boldmath $\sigma$}} = (h,0)$,
${\bf r}_{II} \leftrightarrow {\mbox{\boldmath $\sigma$}} = (-h,0)$,
${\bf r}_{III} \leftrightarrow {\mbox{\boldmath $\sigma$}} = (0,h)$, and
${\bf r}_{IV} \leftrightarrow {\mbox{\boldmath $\sigma$}} = (0,-h)$.
The positions of the neighboring sites are defined such that a hop 
in the appropriate direction leads to ${\bf r}$.  For example,
\begin{equation}
{\bf r} =  {\bf r}_I -
    h \left. \frac{\partial {\bf r}} { \partial \sigma_1}
   \right\vert_{{\bf r}_I} + 
    \left. \frac{h^2}{2} \frac{\partial^2 {\bf r}} {\partial  \sigma_1^2}
   \right\vert_{{\bf r}_I} 
 + O(h^3) \ .
\label{2}
\end{equation}
Note that the $\bf r$
coordinates are considered
to be a fixed function of the $\mbox{\boldmath $\sigma$}$
coordinates: $ {\bf r} = {\bf r} (\mbox{\boldmath $\sigma$})$.
This mapping is independent of time, as the defects
that generate the non-trivial mapping
will be quenched in the two-dimensional crystal.
Inverting eq.\ (\ref{2}) for ${\bf r}_I$ gives
\begin{equation}
{\bf r}_I =  {\bf r} +
    h \left. \frac{\partial {\bf r}} { \partial \sigma_1}
   \right\vert_{{\bf r}} + 
    \left. \frac{h^2}{2} \frac{\partial^2 {\bf r}} { \partial \sigma_1^2}
   \right\vert_{{\bf r}}
 + O(h^3) \ .
\label{3}
\end{equation}
With these expressions for the four neighboring sites, 
eq.\ (\ref{1})  to $O(h)$ becomes
\begin{eqnarray}
\frac {d [V({\bf r}) P({\bf r}, t)]}{d t} &=& \frac{D_0}{h^2} \bigg[
h^2 \frac{\partial^2 r_i}{\partial \sigma_\alpha^2} 
    \frac{\partial (V P)}{\partial r_i}
\nonumber \\ &&
+ h^2
\frac{\partial r_i}{\partial \sigma_\alpha}
\frac{\partial r_j}{\partial \sigma_\alpha}
\frac{\partial^2 (V P)}{\partial r_i \partial r_j}
\bigg] \ ,
\label{4}
\end{eqnarray}
where the summation convention has been used.
Equation (\ref{4}) is exact and leads in the continuum limit
to the general expression for diffusion in curved space \cite{Ikeda}.
The notation 
\begin{equation}
g^{ij} =
\frac{\partial r_i}{\partial \sigma_\alpha}
\frac{\partial r_j}{\partial \sigma_\alpha}
\label{4a}
\end{equation}
will be used.
The shorthand $\partial_i = \partial / \partial r_i$ will also used.
Equation (\ref{4}) is a relation for the probability
distribution in $\bf r$ space.  The relation for the
probability distribution in $\mbox{\boldmath $\sigma$}$ space
requires a Jacobian:
\begin{equation}
 P({\bf r},t) =
G(\mbox{\boldmath $\sigma$}, t) 
 \vert \det \partial \sigma_\alpha / \partial r_i \vert \ .
\label{5b}
\end{equation}
The Jacobian is given by 
\begin{equation}
\vert \det \partial \sigma_\alpha / \partial r_i \vert 
 = \sqrt g({\bf r}) = \left\vert 
\frac{\partial \sigma_x }{\partial r_x}
\frac{\partial \sigma_y}{\partial r_y}
- \frac{\partial \sigma_x }{\partial r_y}
\frac{\partial \sigma_y }{\partial r_x}
\right\vert \ .
\label{6}
\end{equation}
As noted, the Jacobian is 
 $\sqrt g = (\det g_{ij})^{1/2} = 1/(\det g^{ij})^{1/2}$, where
the inverse of the matrix $g^{ij}$ is given by
\begin{equation}
g_{ij} = \frac{\partial \sigma_\alpha}{\partial r_i}
\frac{\partial \sigma_\alpha}{\partial r_j} \ .
\label{4bb}
\end{equation}
The volume of each unit cell is given by $V({\bf r}) = h^2 / \sqrt g({\bf r})$.
By detailed balance, since the rates to hop forward and back between any
two sites on the crystal are the same, the long-time average number of
particles per site must be equal at all sites:
$\lim_{t \to \infty} G(\mbox{\boldmath $\sigma$}, t) = {\rm const}$.
This implies
$\lim_{t \to \infty} P({\bf r}, t) \propto  \sqrt g ({\bf r})$.

Since
\begin{eqnarray}
\partial_i g^{ij} &=& \frac{\partial}{\partial r_i} \left(
\frac{\partial r_i}{\partial \sigma_\alpha}
\frac{\partial r_j}{\partial \sigma_\alpha}
\right)
\nonumber \\ &=&
\frac{\partial r_j}{\partial \sigma_\alpha}
\frac{\partial}{\partial r_i}
\frac{\partial r_i}{\partial \sigma_\alpha}
+
\frac{\partial r_i}{\partial \sigma_\alpha}
\frac{\partial}{\partial r_i}
\frac{\partial r_j}{\partial \sigma_\alpha}
\nonumber \\ 
& =&
\frac{\partial r_j}{\partial \sigma_\alpha}
\frac{\partial}{\partial r_i}
\frac{\partial r_i}{\partial \sigma_\alpha}
+ \frac{\partial^2 r_j}{\partial \sigma_\alpha^2} \ ,
\label{5}
\end{eqnarray}
eq.\ (\ref{4}) becomes
\begin{equation}
\frac{\partial G}{\partial t} = D_0 \left(
 \partial_i g^{ij} -
\frac{\partial r_j}{\partial \sigma_\alpha}
\frac{\partial}{\partial r_i}
\frac{\partial r_i}{\partial \sigma_\alpha}
\right) \partial_j G
+ D_0 g^{ij} \partial_i \partial_j G \ .
\label{5a}
\end{equation}
Finally, given that 
\begin{eqnarray}
\frac{1}{\sqrt g} g^{ij} \partial_i \sqrt g &=& 
-\frac{1}{2} g^{ij} \partial_i \ln \det g^{-1}
\nonumber \\ 
&=& -\frac{1}{2} \frac{\partial r_j}{\partial \sigma_\alpha}
\frac{\partial r_i}{\partial \sigma_\alpha} 
\frac{\partial \ln \det g^{-1}}{\partial r_i}
\nonumber \\ &=&
-\frac{1}{2}
\frac{\partial r_j}{\partial \sigma_\alpha}
\frac{\partial \ln \det g^{-1}}{\partial \sigma_\alpha}
\ ,
\nonumber \\
\label{5aa}
\end{eqnarray}
and
\begin{eqnarray}
\frac{1}{2} 
\frac{\partial \ln \det g^{-1}}{\partial \sigma_\alpha}
&=&
\frac{1}{2} 
g_{ji} \frac{\partial g^{ij}}{\partial \sigma_\alpha}
\nonumber \\ &=&
\frac{1}{2}
\frac{\partial \sigma_\beta}{\partial r_i}
\frac{\partial \sigma_\beta}{\partial r_j}
\frac{\partial}{\partial \sigma_\alpha}
\left(
\frac{\partial r_i}{\partial \sigma_\gamma} 
\frac{\partial r_j}{\partial \sigma_\gamma} 
\right)
\nonumber \\ &=&
\frac{\partial \sigma_\beta}{\partial r_i}
\frac{\partial \sigma_\beta}{\partial r_j}
\frac{\partial r_j}{\partial \sigma_\gamma}
\frac{\partial}{\partial \sigma_\alpha}
\frac{\partial r_i}{\partial \sigma_\gamma}
\nonumber \\ &=&
\frac{\partial \sigma_\beta}{\partial r_i}
\frac{\partial}{\partial \sigma_\alpha}
\frac{\partial r_i}{\partial \sigma_\beta}
\nonumber \\ &=&
\frac{\partial \sigma_\beta}{\partial r_i}
\frac{\partial}{\partial \sigma_\beta}
\frac{\partial r_i}{\partial \sigma_\alpha} ~~~~~({\rm see~below})
\nonumber \\ &=&
\frac{\partial}{\partial r_i}
\frac{\partial r_i}{\partial \sigma_\alpha} \ ,
\label{7a}
\end{eqnarray}
the final expression of the diffusion equation becomes
\begin{equation}
\frac{\partial G}{\partial t} = \frac{1}{\sqrt g}
\partial_i (\sqrt g D_0 g^{ij} \partial_j G) \ .
\label{7}
\end{equation}
This equation applies everywhere except within the cores of topological
defects, because it has been assumed in the second to last line of eq.\ 
(\ref{7a}) that the differentiations commute \cite{SeungII}.
Equation (\ref{7}) is nothing more than the usual
diffusion equation
in curved space, with the familiar Laplace-Beltrami operator
\cite{Kreyszig} replacing the Laplacian of flat space.
The mean-square-displacement is given by
\begin{eqnarray}
\left\langle r^2(t) \right\rangle &= &
\int d {\bf r}  \vert {\bf r} \vert^2 P({\bf r}, t) 
\label{8a}
\\
&= &
\int d {\bf r} \sqrt g({\bf r})  \vert {\bf r} \vert^2 G({\bf r}, t) \ .
\label{8b}
\end{eqnarray}

Equation (\ref{7}) differs from the most general expression for
diffusion in curved space by a term related to the torsion 
\cite{Ikeda}.
  We reevaluate the term $\partial \ln \det g / \partial r_i$:
\begin{eqnarray}
\frac{1}{2} \frac{ \partial \ln \det g}{\partial r_i} &=&
\frac{1}{2} g^{kj} 
\frac{\partial }{\partial r_i}
 g_{jk}
\nonumber \\
&=& \frac{1}{2} 
\frac{ \partial r_k} {\partial \sigma_\alpha}
\frac{ \partial r_j} {\partial \sigma_\alpha}
\frac{\partial }{\partial r_i}
\left(
\frac{ \partial \sigma_\beta}{\partial r_j}
\frac{ \partial \sigma_\beta}{\partial r_k}
\right)
\nonumber \\
&=&
\frac{ \partial r_k} {\partial \sigma_\alpha}
\frac{ \partial r_j} {\partial \sigma_\alpha}
\frac{\partial \sigma_\beta}{\partial r_j}
\frac{\partial }{\partial r_i}
\frac{\partial \sigma_\beta}{\partial r_k}
\nonumber \\
&=&
\frac{ \partial r_k} {\partial \sigma_\alpha}
\frac{\partial }{\partial r_i}
\frac{\partial }{\partial r_k}
\sigma_\alpha
\nonumber \\
&=&
\frac{ \partial } {\partial \sigma_\alpha}
\frac{\partial \sigma_\alpha }{\partial r_i}
\nonumber \\
&&+ 
\frac{ \partial r_k} {\partial \sigma_\alpha}
\left(
\frac{\partial }{\partial r_i}
\frac{\partial }{\partial r_k}
-
\frac{\partial }{\partial r_k}
\frac{\partial }{\partial r_i}
\right)
\sigma_\alpha
\end{eqnarray}
Defining the torsion as
$2 T_{ik}^k = 
( \partial r_k/\partial \sigma_\alpha) (\partial^2/\partial r_i \partial r_k
- \partial^2/\partial r_k \partial r_i) \sigma_\alpha$, note that
\begin{eqnarray}
g^{ij} \frac{1}{2} \frac{ \partial \ln \det g}{\partial r_i} &=&
\frac{\partial r_i} {\partial\sigma_\alpha}
\frac{\partial r_j} {\partial\sigma_\alpha}
\frac{ \partial } {\partial \sigma_\beta}
\frac{\partial \sigma_\beta }{\partial r_i} 
+ 2 g^{ij} T_{ik}^k
\nonumber \\ 
&=&
\frac{\partial r_j} {\partial\sigma_\alpha}
\frac{\partial r_i} {\partial\sigma_\alpha}
\frac{ \partial } {\partial \sigma_\beta}
\frac{\partial \sigma_\beta }{\partial r_i} 
+ 2 g^{ij} T_{ik}^k
\nonumber \\ 
&=&
-
\frac{\partial r_j} {\partial\sigma_\alpha}
\frac{\partial \sigma_\beta }{\partial r_i} 
\frac{ \partial } {\partial \sigma_\beta}
\frac{\partial r_i} {\partial\sigma_\alpha}
+ 2 g^{ij} T_{ik}^k
\nonumber \\ 
&=&
-
\frac{\partial r_j} {\partial\sigma_\alpha}
\frac{ \partial } {\partial r_i}
\frac{\partial r_i}{\partial \sigma_\alpha }
+ 2 g^{ij} T_{ik}^k
\label{8aa}
\end{eqnarray}
Combining eqs.\ (\ref{5a}), (\ref{5aa}),
 and (\ref{8aa}), we find that
the exact expression for the diffusion equation is
\begin{equation}
\frac{\partial G}{\partial t} = \frac{1}{\sqrt g}
\partial_i (\sqrt g D_0 g^{ij} \partial_j G) -
2 D_0 g^{ij}  T_{ik}^k \partial_j G
\ .
\label{8aaa}
\end{equation}
Equation (\ref{8aaa}) is equal to the general
expression for diffusion in curved space \cite{Ikeda}.
The difference between the exact answer, eq.\ (\ref{8aaa}),
 and that assuming that the
order of differentiation commutes, eq.\ (\ref{7}),
 is given by the torsion term.
The torsion is an explicit measure of the non-commutativity
of differentiation and is, therefore, a measure of the defect
density \cite{SeungII}.
  The diffusion equation does not apply within the cores
of defects, where the metric tensor is undefined, and the only
place where the torsion is non-zero.
The effects of the torsion should probably be studied with a
detailed model rather than with 
the long-wavelength, continuum theory of the diffusion equation.
For this reason, we exclude this torsion term
(although see section VII below).
The long range, external to defect core, effects
of the topological defects are, of course, included  in eq.\ (\ref{7}) through
the metric tensor $g^{ij}$ and $\sqrt g$.
Refs.\ \cite{Turski,Turski2} included the torsion term
explicitly, and a series of approximations allowed the
generation of non-physical dynamics.

Equation (\ref{7}) can, alternatively, be derived by consideration of the
field-theoretic representation of the diffusion operator
\cite{Lee1,Lee2}.
In this representation, the Green function is given by
an average over a field:
\begin{equation}
G({\bf r}, t) = \left\langle a({\bf r}, t) \right\rangle
\label{9a}
\ ,
\end{equation}
where the average is taken with respect to the weight $\exp(- S)$.
The particle hopping occurs in $\mbox{\boldmath $\sigma$}$
 space without regard to the
distortion of the crystal, as the rate of hopping is independent of
the distance between lattice sites.
The action for such normal diffusion is given by
\begin{eqnarray}
``S &=& \int_0^\infty dt \int d \mbox{\boldmath $\sigma$} 
\left\{ \bar a [\partial_t + \delta(t)] a
-
D_0
\bar a
 \frac{\partial^2 a}{\partial \sigma_\alpha^2}
\right\}
\nonumber \\
&&
+ \int d \mbox{\boldmath $\sigma$} n_0(\mbox{\boldmath $\sigma$})
        \bar a({\bf r}, 0)\ ,''
\label{9}
\end{eqnarray}
where $n_0$ is the initial density profile, and 
details of the replica indices used to accommodate
averaging over disorder have been suppressed \cite{Kravtsov1,Deem1}.
This action is enclosed in quotations
since the $\mbox{\boldmath $\sigma$}$ space
is not well-defined in the
presence of topological defects.
That the diffusion is normal in $\mbox{\boldmath $\sigma$}$
space, however,
does make it clear that the limiting distribution should be
$\lim_{t \to \infty} G(\mbox{\boldmath $\sigma$}, t) =({\rm const})$.
From eq.\ (\ref{5b}), then, the limiting distribution in ${\bf r}$ space is
given by $\lim_{t \to \infty} P({\bf r}, t) = ({\rm const})  \sqrt g({\bf r})
 = \sqrt g({\bf r}) / \int d {\bf r}' \sqrt g ({\bf r}')$.
While this result may be surprising, note that the defects which distort
the geometry \emph{must}
affect the limiting distribution, unlike the typical case
in differential geometry where the observables are described by a theory
independent of the coordinate system.
This explicit result for the limiting distribution
agrees with the prediction from
the simple detailed balance argument given above. Note that
$\int d {\bf r}  \sqrt g({\bf r})$ is a constant for a given realization of the
quenched disorder.  The long-time normalization factor for the
probability is fixed to be the inverse of this integral
by the initial condition
$P({\bf r},0) = \delta({\bf r})$.  Equation (\ref{7})
for the dynamics conserves $\int d {\bf r} \sqrt g({\bf r}) G({\bf r}, t)$,
hence, the probability distribution,
  $P({\bf r}, t) = \sqrt g({\bf r}) G({\bf r}, t)$,
remains normalized to unity for all times $t \ge 0$.
After change of variables from $\mbox{\boldmath $\sigma$}$ to ${\bf r}$,
again making the assumption of being outside defect cores so
that differentiation commutes, the action becomes
\begin{eqnarray}
S &=& \int_0^\infty dt \int d {\bf r} 
\left\{
\sqrt g
 \bar a [\partial_t + \delta(t)] a
-
\bar a \partial_i[
\sqrt g 
 D_0
g^{ij}
\partial_j a]
\right\}
\nonumber \\
&& + \int d {\bf r} \sqrt g n_0({\bf r}) \bar a({\bf r}, 0) \ .
\label{10}
\end{eqnarray}
Finally, integrating out the $\bar a$ field, using
eq.\ (\ref{9a}), and noting that for the Green function 
$n_0(\mbox{\boldmath $\sigma$}) =
\delta (\mbox{\boldmath $\sigma$})$, the
Fokker-Planck equation is
\begin{equation}
\frac{\partial G}{\partial t} = \frac{1}{\sqrt g}
\partial_i (\sqrt g D_0 g^{ij} \partial_j G) \ ,  
\label{11}
\end{equation}
with $G({\bf r}, 0) = \delta ({\bf r}) / \sqrt g({\bf r})$.
The field-theoretic result, eq.\ (\ref{11}), is the same as
that derived by more physically-motivated means, eq.\ (\ref{7}).

\section{The Model of Topological Disorder}

The topological defects modify the diffusive motion
of the particle by affecting the
$g^{ij}$ in the Fokker-Planck equation.
Once $g^{ij}$ is determined, eqs.\ (\ref{7}) and (\ref{8b})
provide the means to calculate the transport properties.
It is conventional in continuum elasticity theory to relate the
spatial coordinates to the lattice coordinates by
\begin{equation}
{\bf r} (\mbox{\boldmath $\sigma$}) = \mbox{\boldmath $\sigma$} +
{\bf u}({\bf r}) \ ,
\label{12}
\end{equation}
where the displacement field ${\bf u}$ is written in terms of the
${\bf r}$ variables that remain well-defined even in the presence of
topological defects.  The $\mbox{\boldmath $\sigma$}$ space, on the other
hand, does not remain well-defined, since
the effect of disclinations is to add or remove wedges of lattice
sites from $\mbox{\boldmath $\sigma$}$ space,
 and the effect of dislocations is to add or remove half-lines of
lattice sites from $\mbox{\boldmath $\sigma$}$ space.
  For a dislocation at the origin with
Burgers vector {\bf b}, the displacement fields are given by
\cite{Nabarro}
\begin{eqnarray}
2\pi u_i^{\rm disloc} &=& 
-\frac{(\mu+\lambda)}{(2\mu+\lambda)} 
 \frac{\epsilon_{kl}b_l r_i r_k}{r^2} +
 b_i \tan^{-1} \frac{r_y}{r_x}
\nonumber \\ && 
-
\frac{\mu}{2\mu+\lambda} 
\epsilon_{li} b_l
\ln{\frac{r}{h}} \ ,
\label{13}
\end{eqnarray}
where $\mu$ and  $\lambda$ are the two-dimensional Lam{\'e} coefficients,
$\epsilon_{11} = \epsilon_{22} = 0$, and
$\epsilon_{12} = -\epsilon_{21} = 1$.
Similarly, for a disclination of strength $s$ at the origin, the
displacement fields are given by
\begin{eqnarray}
2 \pi u^{\rm disclin}_i &=& 
- \frac{(\mu+\lambda) }    {2(2\mu+ \lambda)}  s r_i
- s \epsilon_{ik}r_k \tan^{-1}{\frac{r_y}{r_x}}
\nonumber \\ && 
+ \frac{\mu }{2\mu+\lambda} 
 s r_i \ln(r/h) \ .
\label{14}
\end{eqnarray}
Equation (\ref{14}) differs from the simplified distortion field used
in \cite{Turski} by the inclusion of the strain field representing the local
lattice contraction and expansion.  These are the terms in eq. (\ref{14})
that depend on the Lam{\'e} coefficients.
Since linear elasticity theory is used, the dislocation field is given by the
dipole limit of two superimposed disclination fields:
\begin{equation}
u^{\rm disloc}_i = (b_l/s)
\epsilon_{jl} \partial_j u^{\rm
disclin}_i + {\rm const} \ .
\label{14a}
\end{equation}
The derivatives of the displacement fields are required to evaluate
$g^{ij}$ from eq.\ (\ref{4a}).  The dislocation
fields are preferable for this calculation, as
 they lead to well-defined Fourier transforms:
\begin{eqnarray}
\hat {\partial_x u_x} \hbox{}^{\rm disloc} &=&
 i \left[
\frac{\mu + \lambda}{2 \mu + \lambda}
\frac{2 k_x^2 k_y}{k^4}
 - \frac{k_y}{k^2}
\right] \hat b_x ({\bf k})
\nonumber \\ &&
+ i \left[ 
\frac{\mu + \lambda}{2 \mu + \lambda}
\frac{k_x(-k_x^2 + k_y^2)}{k^4} +
\frac{\mu}{2 \mu + \lambda}
\frac{k_x}{k^2}
\right] \hat b_y ({\bf k})
\nonumber \\ 
\hat {\partial_y u_x} \hbox{}^{\rm disloc} &=&
 i \left[
\frac{\mu + \lambda}{2 \mu + \lambda}
\frac{2 k_x k_y^2}{k^4}
 + \frac{k_x}{k^2}
\right] \hat b_x ({\bf k})
\nonumber \\ &&
+ i \left[ 
\frac{\mu + \lambda}{2 \mu + \lambda}
\frac{k_y(-k_x^2 + k_y^2)}{k^4} +
\frac{\mu}{2 \mu + \lambda}
\frac{k_y}{k^2}
\right] \hat b_y ({\bf k})
\nonumber \\ 
\hat {\partial_x u_y} \hbox{}^{\rm disloc} &=&
i \left[ 
\frac{\mu + \lambda}{2 \mu + \lambda}
\frac{k_x (-k_x^2 + k_y^2)}{k^4}
- \frac{\mu}{2 \mu + \lambda}
 \frac{k_x}{k^2}
\right] \hat b_x ({\bf k})
\nonumber \\ &&
+ i \left[ 
-\frac{\mu + \lambda}{2 \mu + \lambda}
\frac{2 k_x^2 k_y}{k^4} -
\frac{k_y}{k^2}
\right] \hat b_y ({\bf k})
\nonumber \\ 
\hat {\partial_y u_y} \hbox{}^{\rm disloc} &=&
i \left[  
\frac{\mu + \lambda}{2 \mu + \lambda}
\frac{k_y (-k_x^2 + k_y^2)}{k^4}
- \frac{\mu}{2 \mu + \lambda}
 \frac{k_y}{k^2}
\right] \hat b_x ({\bf k})
\nonumber \\ &&
+ i \left[
-\frac{\mu + \lambda}{2 \mu + \lambda}
\frac{2 k_x k_y^2}{k^4} +
\frac{k_x}{k^2}
\right] \hat b_y ({\bf k}) \ .
\label{15}
\end{eqnarray}
Note that the $x$ and $y$
derivatives of the strain fields are not simply related by
the ratio $k_x/k_y$, due to the presence of the defects.
The linearity of elasticity theory has been used to
accommodate a density field of defects with Burgers vectors
given by ${\bf b}({\bf r})$.

The dislocations are assumed to be 
distributed randomly in the material with correlation function
\begin{equation}
\left\langle \hat b_i({\bf k}_1) \hat b_j({\bf k}_2) \right\rangle
= \delta_{ij} (2 \pi)^2 \delta({\bf k}_1 + {\bf k}_2)
\hat \chi({\bf k}_1 + {\bf k}_2) \ ,
\label{16}
\end{equation}
where
\begin{equation}
\hat \chi({\bf k}) = \gamma k^n \exp(-\beta k^2) \ .
\label{16a}
\end{equation}
Physically, we expect this model of dislocations
should generate identical dynamics to one in which
disclinations are randomly distributed
with correlation function
\begin{equation}
\left\langle \hat s({\bf k}_1) \hat s({\bf k}_2) \right\rangle
= (2 \pi)^2 \delta({\bf k}_1 + {\bf k}_2)
\vert {\bf k}_1 + {\bf k}_2 \vert^2 \hat \chi({\bf k}_1 + {\bf k}_2) \ .
\label{17}
\end{equation}
This physical expectation is a mathematical consequence of
eq.\ (\ref{14a}).

With these results in hand, we are now in a position to
calculate the action for the  field theoretic description of
the Green function.  The terms in eq.\ (\ref{10})
are expressed to linear and quadratic order in $u^{\rm disloc}$, and then an
average over the random distribution of dislocations is taken.
  In fact, since  eq.\ (\ref{8a}) is preferable to
eq.\ (\ref{8b}),  the theory is written in terms of the
fields $\bar a, c$, where $c = \sqrt g a$, and $P = \langle c \rangle$.
The action is
\begin{eqnarray}
S &=& \int_0^\infty dt \int d {\bf r}
 \bar a [\partial_t 
- ( D_0 + \delta D_{II}) \nabla^2
+ \delta(t)] c
\nonumber \\
&& + \int d {\bf r} \delta({\bf r}) \bar a({\bf r}, 0)
+ S_I \ ,
\label{18}
\end{eqnarray}
where
\begin{eqnarray}
S_I &=& -2 D^2
\int d t_1 d t_2 \int_{{\bf k}_1 {\bf k}_2 {\bf k}_3 {\bf k}_4}
\nonumber \\ &&
 \times (2 \pi)^2 \delta({\bf k}_1+{\bf k}_2+{\bf k}_3+{\bf k}_4)
\nonumber \\ &&\times
\hat{\bar a}({\bf k}_1, t_1)
\hat{     c}({\bf k}_2, t_1)
\hat{\bar a}({\bf k}_3, t_2)
\hat{     c}({\bf k}_4, t_2)
\nonumber \\ && \times
\bigg[
\frac{\mu}{2 \mu + \lambda} k_1^2
 +
\frac{\mu + \lambda}{2 \mu + \lambda}
\frac{{\bf k}_1 \cdot {\bf k}_2 (k_1^2  + k_2^2) + 2 k_1^2 k_2^2}
{\vert {\bf k}_1 + {\bf k}_2 \vert^2}
\bigg]
\nonumber \\ && \times
\bigg[
\frac{\mu}{2 \mu + \lambda} k_3^2
 +
\frac{\mu + \lambda}{2 \mu + \lambda}
\frac{{\bf k}_3 \cdot {\bf k}_4 (k_3^2  + k_4^2) + 2 k_3^2 k_4^2}
{\vert {\bf k}_3 + {\bf k}_4 \vert^2}
\bigg]
\nonumber \\ && \times
\frac{\hat \chi({\bf k}_1 + {\bf k}_2)}
{\vert {\bf k}_1 + {\bf k}_2 \vert^{2}}
 \ ,
\label{20}
\end{eqnarray}
where the notation
$\int_{\bf k}$ stands for $\int d^2 {\bf k} / (2 \pi)^2$.
The term resulting from a non-zero average of $(u^{\rm disloc})^2$ is
\begin{eqnarray}
\delta D_{II} &=&  
\frac{\gamma D_0}{2 \pi} 
\left[
 \frac{13 \lambda^2 + 16 \mu \lambda + 25 \mu^2}
{9 (2 \mu + \lambda)^2}
\right]
\nonumber \\ && \times
\int_0^\infty  dk~
k^{n-1} \exp(-\beta k^2) 
\nonumber \\
&=& 
\frac{\gamma D_0}{2 \pi}
\frac{\Gamma(n/2) }{ 2 \beta^{n/2}}
\left[
 \frac{13 \lambda^2 + 16 \mu \lambda + 25 \mu^2}
{9 (2 \mu + \lambda)^2}
\right]
\ .
\nonumber \\
\label{21}
\end{eqnarray}
Exactly the same theory is generated if the
correlation function eq.\ (\ref{17}) is used
with the disclination
displacements given by eq.\ (\ref{14}).

\section{Topological Disorder Reduces the Diffusion Constant}

For the model with $n>0$, the topological disorder reduces the diffusion
coefficient by a finite amount.  The finite contribution of $\delta D_{II}$
is explicit in eq.\ (\ref{21}).  Moreover, standard power counting
arguments \cite{Justin} show that 
non-perturbative, renormalization effects can be expected
from eq.\ (\ref{20}) only for $n \le 0$.
From perturbation theory on eq.\ (\ref{20}) for $n>0$, 
the contribution to the diffusion coefficient is found to be
\begin{equation}
\delta D_I = 
-\frac{\gamma D_0}{2 \pi} 
\frac{\Gamma(n/2) }{ 2 \beta^{n/2}}
\left[
\frac{4 \mu^2
 + 2 (\mu + \lambda)^2}{(2 \mu + \lambda)^2}
\right] \ .
\end{equation}
The total contribution to the diffusion coefficient is, therefore,
\begin{equation}
\delta D = 
-\frac{\gamma D_0}{2 \pi} 
\frac{\Gamma(n/2) }{ 2 \beta^{n/2}}
\left[
\frac{29 \mu^2 + 20 \mu \lambda + 5 \lambda^2}
{9 (2 \mu + \lambda)^2}
\right] \ .
\label{22}
\end{equation}

To demonstrate the behavior of this model, we perform numerical
simulations.  The dislocation density fields are constructed 
with correlation function eq.\ (\ref{16}) for $n=2$ using
the method of ref.\ \cite{Victor}. Equation
(\ref{15}) and an inverse fast Fourier transform are used to
calculate the displacement fields in real space.
The matrix $g^{ij}$ is calculated as the inverse of
the matrix $g_{ij}$ given by eq.\ (\ref{4bb}),
and the relation 
$\mbox{\boldmath $\sigma$} = {\bf r} - {\bf u}({\bf r})$ 
is used.

The Fokker-Planck equation, eq.\ (\ref{7}),
can be considered to result from many small hops,
the net effect of which is Gaussian, diffusive
motion.  So that a hopping process on a lattice
reproduces this differential equation, 
the average and mean-square displacements
must be correct at each lattice site.
Interestingly, this differential equation can be evaluated by Monte Carlo
methods  on a
perfect, square lattice, even though the differential equation
itself describes the motion of a particle in a distorted
geometry.  To first order in the time step, the mean displacement is given by
\begin{eqnarray}
\left\langle r_i (\Delta t) \right\rangle
&=& \int d {\bf r} \sqrt g~ r_i G({\bf r}, \Delta t)
\nonumber \\
& =& \int_0^{\Delta t} dt \int d {\bf r} \sqrt g~ r_i 
\partial_t G({\bf r}, t)
\nonumber \\ &&
 + \int d {\bf r} \sqrt g~ r_i G({\bf r}, 0)
\nonumber \\
&=& \frac{D_0 \Delta t}{\sqrt g} \partial_j ( \sqrt g g^{ij})
\ , 
\label{22a}
\end{eqnarray}
where eq.\ (\ref{7}) and integration by parts twice has been used in the
last step.
Similarly, to first order, the mean-square displacement is given by
\begin{eqnarray}
\left\langle r_i (\Delta t) 
r_j(\Delta t) \right\rangle
&=&
 \int d {\bf r} \sqrt g~ r_i r_j G({\bf r}, \Delta t)
\nonumber \\
& =& \int_0^{\Delta t} dt \int d {\bf r} \sqrt g~ r_i r_j
\partial_t G({\bf r}, t)
\nonumber \\ &&
 +  \int d {\bf r} \sqrt g~ r_i r_j G({\bf r}, 0)
\nonumber \\
&=& 2 D_0 \Delta t g^{ij}
\ , 
\label{22b}
\end{eqnarray}
where eq.\ (\ref{7}) and integration by parts twice 
has again been used in the last step.

Eight hopping rates are defined, consistent with
the specifications of eqs.\ (\ref{22a}--\ref{22b}).
So that the non-diagonal terms of $g^{ij}$ are properly
reproduced, both nearest- and next-nearest-neighbor hops are required.
The rate for each hopping event is
\begin{eqnarray}
T_i({\bf r} \to {\bf r}+ \Delta {\bf r})
 = \frac{D_0}{h^2} \left[ \frac{g({\bf r}+
 \Delta {\bf r})}{g({\bf r})}\right]^{1/4}
\nonumber \\
\times \frac{1}{2}
\left[ f({\bf r}) + f({\bf r}+ \Delta {\bf r})
\right] \ .
\label{23}
\end{eqnarray}
The function $f$ is given by
\begin{eqnarray}
f &=& g^{11} - \epsilon, {\rm~for~} \Delta {\bf r} = (\pm h, 0)
\nonumber \\
f &=& g^{22} - \epsilon, {\rm~for~} \Delta {\bf r} = (0, \pm h)
\nonumber \\
f &=& (g^{12} + \epsilon)/2, {\rm~for~} \Delta {\bf r} = (\pm h, \pm h)
\nonumber \\
f &=& (-g^{12} + \epsilon)/2, {\rm~for~} \Delta {\bf r} = (\pm h, \mp h) \ ,
\label{24}
\end{eqnarray}
with $\epsilon = \vert g^{12}\vert$.
The transition rates in eq.\ (\ref{23}) explicitly satisfy detailed
balance for the equilibrium distribution 
$\lim_{t \to \infty} P({\bf r}, t) = (\rm const) \sqrt g({\bf r})$.
These rates give the correct average and mean-square
displacements to $O(h)$, eqs.\ (\ref{22a}--\ref{22b}), when
$\Delta t = 1/(\sum_i T_i)$.  
These results imply that
the Monte Carlo procedure evaluates the differential eq.\
(\ref{4}), and so
eq.\ (\ref{8a}) can be used to calculate the mean-square-displacements.
The procedure of ref.\ \cite{Victor} is used
to perform the simulation of this random process, where the
particle is moved to one of the neighboring eight sites with probability
$\Delta t T_i$, and time is
incremented by $dt = -\Delta t \ln (x)$, where $x$ is a
uniform random number, $0 < x \le 1$.

\begin{figure}[tbp]
\centering
\leavevmode
\psfig{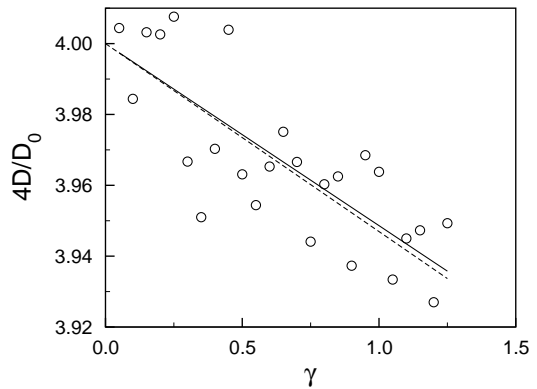}
\caption{Shown are simulation results for the reduction in the diffusion
coefficient for the case $n=2$, $\mu = \lambda$, and $\beta = 4$.
The error bars are roughly $\pm 0.01$.
The best linear fit to the simulation data is shown (solid line).
The simulation data are compared to  perturbation theory (dashed line),
eq.\ \ref{22}, $\delta D/D_0 = -\gamma / (6 \pi \beta)$.
}
\label{fig2}
\end{figure}

The results of the numerical simulations are shown in Fig.\ \ref{fig2}.
The calculations were performed for the case $\mu = \lambda$, $n=2$, 
$h=1$, and
$\beta  = 4$.  The simulations were done on $4096 \times 4096$ lattices
for a total of 500000 steps and averaged over 100000 particles.
The strength of the disorder was varied between $0 < \gamma < 1.25$.
For larger values of $\gamma$, the transition rates
specified by eqs.\ (\ref{23}--\ref{24}) became negative at some
of the lattice sites.
Also shown is a fit to the functional form $\delta D/D_0 = 1 - a x$.
The fit to the simulation data of
$a = 0.01285 \pm 0.0051$
 is in excellent agreement with the theoretical
of result $a = 0.01326$ from eq.\  (\ref{22}).

\section{Possible Anomalous Diffusion}

The case $n=0$ is interesting, as perturbation theory
for the diffusion coefficient formally diverges.
While this theory has the same upper critical dimension,
$d_c = 2$,
 as the problem of diffusion of an ion in the electrostatic field of
random, quenched charges \cite{Bouchaud}, the
interaction term, eq.\ (\ref{20}), is quite different.
In comparison to the analogous term for diffusion in the 
random potential (\emph{e.g.}\ term $S_3$ of ref.\ \cite{Deem1} with
$\hat \chi_{vv}(k) = \hat \chi(k) / k^2$),
the term proportional to $\mu$ is new, as are the
factors $-k_1^2 + 2 [k_1^2 k_2^2 - ({\bf k}_1 \cdot {\bf k}_2)^2]/
\vert {\bf k}_1 + {\bf k}_2 \vert^2$
 in the term proportional to $\mu + \lambda$.
Indeed, as we will see, the present interaction term is more
difficult to analyze than is the analogous one from diffusion in a random
potential.
Formally, the case of $n \le 0$ leads to large distortions of the lattice
for arbitrarily small $\gamma$, which implies that the assumption of
linear elasticity used to calculate the
strain fields breaks down.  We can, however, treat the
dynamical behavior implied by eqs.\ (\ref{18}--\ref{20}) as an
interesting mathematical question.
A technical detail is that we supplement the correlation function
eq.\ (\ref{16a}) with the condition $\hat \chi({\bf 0}) \equiv 0$ so that
the displacement fields of eq.\ (\ref{15}) are well-defined for
$k=0$.  This suppresses macroscopic size fluctuations of
the sample.

Before applying renormalization group theory, the
terms in the field theory must be known.
The quartic interaction term, eq.\ (\ref{20}), is known.
The contribution to the propagator, eq.\ (\ref{21}), while 
explicit, leads
to a formal divergence of the short-time diffusion
coefficient.  Numerical simulations show that the
local diffusivity tensor, $D_0 g^{ij}$, can be large but is
never vanishingly small.  The locations of
large local diffusivity, moreover, are isolated.  The apparent
divergence of $\delta D_{II}$ is, thus, simply the result of
particles rapidly hopping away from a few isolated locations.
These physical considerations suggest
that the divergence of  $\delta D_{II}$ is washed out by
spatial averaging and is not important for the long-time dynamics.
We can, therefore, assume a finite local diffusivity.
Numerical simulations of the dynamics, to be described below,
bear out this assumption of a finite short-time diffusivity.
Indeed, a finite short-time diffusivity is assured for finite lattice
sizes by the elimination of the $\hat \chi ({\bf 0})$ mode.
The anomalous dynamics, then, is observed on finite lattices for
time scales that are less than the characteristic time it takes to
travel across the lattice.

\begin{figure}[tbp]
\centering
\leavevmode
\psfig{file=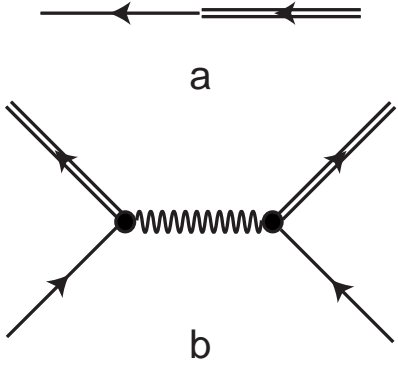,height=2in,clip=}
\caption{
a) Diagram representing the propagator. The arrow points
in the direction of increasing time, and double lines represent the
bar fields.
b) Disorder vertex $\gamma$.
}
\label{fig3}
\end{figure}

\begin{figure}[tbp]
\centering
\leavevmode
\psfig{file=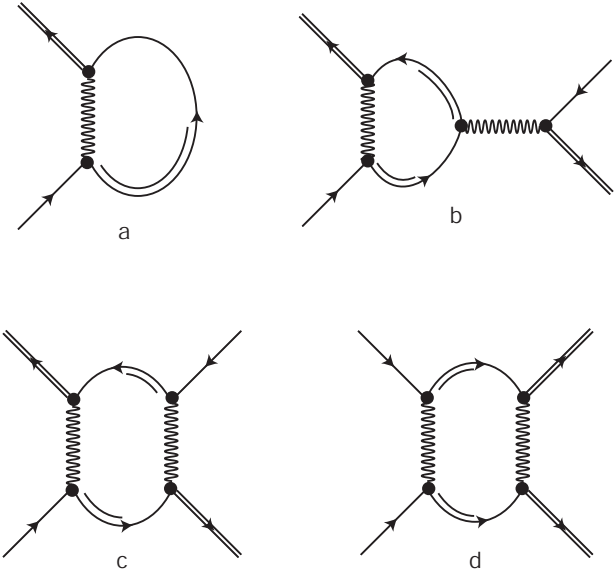,height=3in,clip=}
\caption{
One-loop diagrams: a) self-energy diagrams contributing to $D$.
b,c,d) vertex diagrams contributing to $\gamma$.
Diagrams (c) and (d) cancel.
}
\label{fig4}
\end{figure}

We apply renormalization group theory to the action (\ref{18}--\ref{20}) to
take into account the effects of nonzero $\gamma$.
To one-loop order, self-energy and vertex diagrams are summarized
in Figs.\ \ref{fig3} and \ref{fig4}.
The flow equations are integrated to a time small enough so that
perturbation theory applies.
In this regime, matching theory is used to determine
the constants of integration for the flow equations.
Momenta in the range $\Lambda/b < k < \Lambda$ are
integrated over, and the fields are rescaled by
$\hat{\bar a}'(b {\bf k},b^{-z} t) =
\hat{\bar a}( {\bf k},t) / \bar \alpha$
 and
$\hat { c}'(b {\bf k},b^{-z} t) =
\hat { c}( {\bf k},t) /  \alpha$.
The relations $\alpha =1, \bar \alpha = b^2$ are used to
achieve a fixed point and to keep the time derivative in $S$
constant.  The flow parameter is defined by $l = \ln b$.
 We determine the
dynamical exponent, $z$, by requiring that the diffusion
coefficient remain unchanged. 
Defining 
$\gamma_1^2 = \gamma \mu^2 / (2 \mu + \lambda)^2$,
$\gamma_2^2 = \gamma \mu (\mu+ \lambda) / [2(2 \mu + \lambda)^2]$, and
$\gamma_3^2  = \gamma (\mu +\lambda)^2 / [4 (2 \mu + \lambda)^2]$,
the contributions to the parameters from the one-loop diagrams of
fig.\ \ref{fig4}  are
\begin{eqnarray}
\frac{d \ln D}{dl} &=&
z -2 
- \frac{2} {\pi} (\gamma_1^2 + 2 \gamma_3^2)
\nonumber \\
\frac{d \ln \gamma_1^2 }{d l} &=& 
2(z-2)
- \frac{4} {\pi} (\gamma_1^2 + 2 \gamma_3^2)
\nonumber \\
\frac{d \ln \gamma_2^2 }{d l} &=& 
2(z-2)
- \frac{2} {\pi} (\gamma_1^2 + 3 \gamma_3^2 - 2 \gamma_1 \gamma_3)
\nonumber \\
\frac{d \ln \gamma_3^2 }{d l} &=&
2(z-2)
- \frac{4} {\pi} (\gamma_3^2 - 2 \gamma_1 \gamma_3)
 \ .
\label{25a}
\end{eqnarray}
From the requirement that the diffusion coefficient remain fixed,
the dynamical exponent is
\begin{equation}
z = 2 
+ \frac{2} {\pi} (\gamma_1^2 + 2 \gamma_3^2) \ .
\label{26}
\end{equation}
Using eq.\ (\ref{26}) in eq.\ (\ref{25a}), the flow equations become
\begin{eqnarray}
\frac{d \ln \gamma_1^2 }{d l} &=& 
0
\nonumber \\
\frac{d \ln \gamma_2^2 }{d l} &=& 
\frac{2 }{\pi} 
   ( \gamma_1 +  \gamma_3)^2
\nonumber \\
\frac{d \ln \gamma_3^2 }{d l} &=&
 \frac{4 }{\pi}
    ( \gamma_1 +  \gamma_3)^2
 \ .
\label{25}
\end{eqnarray}
As expected, the flow equations show that there are only two 
independent parameters, $\gamma_1$ and $\gamma_3$, resulting from
renormalization of the two Lam{\'e} coefficients.  In other words,
the relation $\gamma_2^2(l) = \gamma_1(l) \gamma_3(l)$ is maintained
under the renormalization.

Unexpectedly, however, the flow equations show that the $\gamma_2(l)$
and $\gamma_3(l)$ are growing.  
Indeed, these one-loop flow equations predict
$\gamma_3(l)$ flows to infinity at a finite time corresponding to
$l = [\pi / (2 \gamma_1^0)^2]\{
\ln [( \gamma_1^0 +  \gamma_3^0 )/ \gamma_3^0]
-  \gamma_1^0/( \gamma_1^0 +  \gamma_3^0 )
\}
$.
The divergence of this parameter implies that higher order terms must
be kept in the flow equation to derive a controlled result.
It may also be the case that 
terms higher order in $u^{\rm disloc}$ must be
kept in the expansion of the action (\ref{10}).  

If the renormalization of the parameters is assumed to be
controlled by higher-loop corrections and small,
the dynamical exponent can be used to determine the scaling 
exponent for the mean-square displacement at long times: 
\begin{equation}
\left\langle r^2(t) \right\rangle \sim ({\rm const}) t^{1- \delta} \ .
\label{26a}
\end{equation}
The renormalized time flows as
\begin{equation}
t(l^*) = t e^{-\int_0^{l^*} z(l) dl } = t_0 \ ,
\label{27}
\end{equation}
where the flow equations are stopped at $l^*$ so that
$t_0 \approx h^2 /(4 D_0)$.
The renormalized mean-square displacement flows as
\begin{equation}
\left\langle r^2(t) \right\rangle =
e^{2 l} \left\langle r^2[t(l), l] \right\rangle \ .
\label{28}
\end{equation}
Finally, at the matching
\begin{equation}
\left\langle r^2[t(l^*), l^*] \right\rangle = 4 D t(l^*) \ ,
\label{29}
\end{equation}
since the time is short enough so that the disorder does not
significantly affect the motion of the particle.
In other words, it is assumed that at short times
the diffusion coefficient remains finite, despite the formal appearance
of $\delta D_{II}$ in eq.\ (\ref{21}).
Putting these matching results together with the dynamical
exponent, the mean-square displacement is found to scale at long times as
\begin{equation}
\left\langle r^2(t) \right\rangle \sim 
   ({\rm const}) t^{1/[1+(\gamma_1^2 + 2 \gamma_3^2)/\pi]} \ .
\label{30}
\end{equation}

To test whether anomalous scaling occurs in the full non-linear model, we 
perform numerical simulations.  The transition rates
from eqs.\ (\ref{23}--\ref{24}) cannot be used, as they are
negative even for small values of $\gamma$.
We, therefore, develop a new strategy based upon the idea that
diffusion locally follows a Gaussian probability
distribution with mean and variance specified by
eqs.\ (\ref{22a}--\ref{22b}).
The time increment $\Delta t$ is chosen
so that $\max (\vert \langle \delta r_i \rangle \vert,
\vert \langle \delta r_i \delta r_j \rangle \vert)$ is on the order of unity.
This is done by choosing $1/\Delta t$ to be the maximum of the 
absolute values of the 
two average displacements in eq.\ (\ref{22a}) and the two
eigenvalues of the matrix $D_0 g^{ij}$.
Defining the matrix $w = ( 2 D_0 \Delta t g^{ij})^{1/2}$, the random
displacements of the diffusing particle are given by the relations
\begin{eqnarray}
\Delta x &=& \frac{D_0 \Delta t }{\sqrt g} \partial_j (\sqrt g g^{1j})
+ w^{11} z_1 + w^{12} z_2
\nonumber \\
\Delta y &=& \frac{D_0 \Delta t }{\sqrt g} \partial_j (\sqrt g g^{2j})
+ w^{21} z_1 + w^{22} z_2 \ ,
\label{31}
\end{eqnarray}
where $z_1$ and $z_2$ are independent, Gaussian random variables with
zero mean and unit variance.
This approach reproduces the Fokker-Planck equation (\ref{4})
in the limit of a small lattice spacing and time
increment.  For a finite
lattice spacing, the diffusion coefficient,
and possibly the scaling exponent $\delta$,
contain discretization errors.

\begin{figure}[btp]
\centering
\leavevmode
\psfig{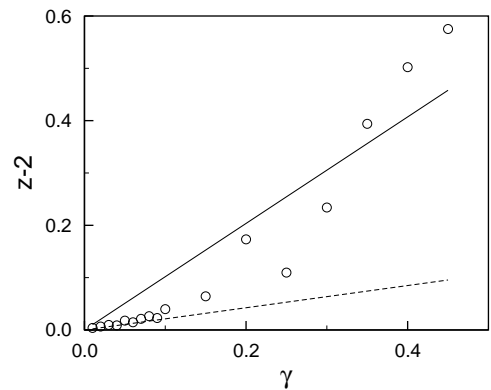}
\caption{Shown are simulation results for the scaling exponent
for the case $n=0$, $\mu = \lambda$, and $\beta = 4$.
The error bars are given roughly by the scatter in the data.
The best linear fit to the simulation data is shown (solid line).
 Also shown (dashed line)
is the prediction assuming that the other parameters
do not flow,
$z = 2 + 2 \gamma / (3 \pi)$.
}
\label{fig5}
\end{figure}

The results of numerical simulations with this
scheme are shown in Fig.\ \ref{fig5}.
The calculations were performed for the case $\mu = \lambda$, $n=0$, 
$h=1$, and
$\beta  = 4$.  The simulations were done on $4096 \times 4096$ lattices
for a total of 1000000 steps and averaged over 100000 particles.
Also shown is a fit to the functional form of eq.\ (\ref{26a}).
The simulation results are approximately
 fit by $(z- 2)/\gamma =  1.02 \pm 0.28$.
If it is assumed that
none of the parameters flow, the scaling exponent 
is given by $(z-2)/\gamma = 2/(3 \pi) \approx 0.2122$.
By comparison, the simulation results suggest that there is
substantial positive renormalization of the parameters, as
is suggested by eq.\ (\ref{25}).  

\section{Discussion}

A Fokker-Planck equation for diffusion on the surface of
a crystal with topological defects, eq.\ (\ref{7}), has been derived
by two independent methods.  As expected, the usual 
diffusion equation in curved space is derived.
An additional assumption
of $\sqrt g = 1$ of previous, approximate treatments \cite{Turski} has also
been removed in the present calculation through the use of the exact
strain field eq.\ (\ref{14}).
The theory of random dislocations is shown to be equivalent to a
theory of random disclinations, where a simple factor of $k^2$ 
relates the correlation functions of the two models of disorder,
eqs.\ (\ref{16}) and (\ref{17}).

The field theory for disorder, eqs.\ (\ref{18}--\ref{20}), is 
explicitly shown to be distinct from that for diffusion of an ion
in a random electrostatic potential field.
One consequence of this difference is that the renormalization group
flow equations are more involved to analyze, with 
one-loop results unable to render a controlled
prediction.

Topological disorder slows down a diffusing particle, as
shown by eq.\ (\ref{22}).
This reduced transport should be observable on the surfaces of crystals with
quenched disclination or dislocation defects.
While the effect is subtle, it would be an interesting one to observe
experimentally.  The present computer simulation results suggest
such observations should be feasible.

For singular disorder, $n \le 0$ in two dimensions,
the model of topological disorder leads to subdiffusive motion of the
particle.  Of course, for such singular
disorder, the assumption of linear
elasticity breaks down.  
Moreover, the energy of a distribution of topological
defects with net dipole moment
becomes super-extensive due to large strain fields at the
edges of the two-dimensional crystal \cite{SeungII}.
Nonetheless, the suggestion that subdiffusion is the
mathematical result of motion in the, possibly approximate,
 random displacement fields of linear elasticity
theory is interesting.  Renormalization group
arguments are suggestive of such subdiffusion, although
one-loop results are unable to capture the exponents quantitatively.

Numerical simulations accurate to all orders in the displacement fields
suggest that the motion is, indeed, subdiffusive.
These numerical simulations suggest that there is significant renormalization
of the disorder strength parameter, in contrast to the case of diffusion in 
random potential fields \cite{Bouchaud}. 
Interestingly, the renormalization of
$z$ appears less significant for smaller values of $\gamma$, although this may
be because the crossover time for renormalization is large for 
small $\gamma$ and longer than the observed simulation time.
These simulations suggest a
power law behavior of the mean square displacement, although
localization at exceptionally long times cannot be ruled out,
in principle.

\section{What if Torsion is Included?}
We here comment on the impact of the torsion term within
the continuum theory of the diffusion equation.  The torsion 
term is  evaluated as
\begin{eqnarray}
2 T_{ik}^k \equiv T_i &=&
\left( \frac{\partial r_k}{\partial \sigma_\alpha} \right)
\left(
\frac{\partial^2}{\partial r_i \partial r_k}
- \frac{\partial^2}{\partial r_k \partial r_i}
\right) \sigma_\alpha
\nonumber \\ 
&=& -\epsilon_{ij} 
\frac{\partial r_j}{\partial \sigma_\alpha} b_\alpha \ .
\label{31aa}
\end{eqnarray}
Expanding eq.\ (\ref{8aaa}) to linear order in $u^{\rm disloc}$,
we find that the interaction term, previously eq.\ (\ref{20}), becomes
\begin{eqnarray}
S_I &=&  \frac{D^2}{2}
\int d t_1 d t_2 \int_{{\bf k}_1 {\bf k}_2 {\bf k}_3 {\bf k}_4}
\nonumber \\ &&
 \times (2 \pi)^2 \delta({\bf k}_1+{\bf k}_2+{\bf k}_3+{\bf k}_4)
\nonumber \\ &&\times
\hat{\bar a}({\bf k}_1, t_1)
\hat{     c}({\bf k}_2, t_1)
\hat{\bar a}({\bf k}_3, t_2)
\hat{     c}({\bf k}_4, t_2)
\nonumber \\ && \times
\bigg[
\left( 
2 \gamma_1 \frac{k_1^2} {\vert {\bf k}_1 + {\bf k}_2 \vert^{2}}
 +
4 \gamma_3 
\frac{{\bf k}_1 \cdot {\bf k}_2 (k_1^2  + k_2^2) + 2 k_1^2 k_2^2}
{\vert {\bf k}_1 + {\bf k}_2 \vert^4}
\right)
\nonumber \\ && \times
({\bf k}_1 + {\bf k}_2) + \sqrt \gamma {\bf k}_2
\bigg]
\nonumber \\ && \cdot
\bigg[
\left(
2 \gamma_1 
\frac{k_3^2} {\vert {\bf k}_3 + {\bf k}_4 \vert^{2}}
 +
4 \gamma_3 
\frac{{\bf k}_3 \cdot {\bf k}_4 (k_3^2  + k_4^2) + 2 k_3^2 k_4^2}
{\vert {\bf k}_3 + {\bf k}_4 \vert^4}
\right)
\nonumber \\ && \times
({\bf k}_3 + {\bf k}_4) + \sqrt \gamma {\bf k}_4
\bigg]
\nonumber \\ && \times
\hat \chi({\bf k}_1 + {\bf k}_2)
 \ .
\label{32}
\end{eqnarray}
Exactly the same theory is generated if the
correlation function eq.\ (\ref{17}) is used
with the disclination
displacements given by eq.\ (\ref{14}).
The inclusion of the torsion term has generated the additional
terms proportional to $\sqrt \gamma k_2$ and 
$\sqrt \gamma k_4$.  Applying perturbation theory to $S_I$,
we find that a mass term is generated,
$\delta m = - 2 D_0 \gamma_1 \sqrt \gamma $.
This term is exactly canceled by a mass term arising
from the average of terms proportional to
$(u^{\rm disloc})^2$, which must be the case since the
master equation (\ref{1}) conserves probability.
No contribution to the diffusivity is generated
by the average average of terms proportional to
$(u^{\rm disloc})^2$.  From the average of $S_I$, we
find an additional negative contribution
to the diffusivity:
\begin{equation}
\delta D_{III} = 
-\frac{\gamma D_0}{2 \pi} 
\frac{\Gamma(n/2) }{ 2 \beta^{n/2}}
\left[
\frac{2 (\mu + \lambda)}{(2 \mu + \lambda)}
\right] \ .
\label{33}
\end{equation}
Within the approximation of the continuum diffusion equation, then, the
torsion term generates an additional contribution to the effective
diffusivity when $n>0$.  Note that this contribution is a result of
correlated drift terms that exist solely within the cores of the defects.
There is no reason to expect that this contribution is universal or
even well-described by continuum theory.

\begin{figure}[btp]
\centering
\leavevmode
\psfig{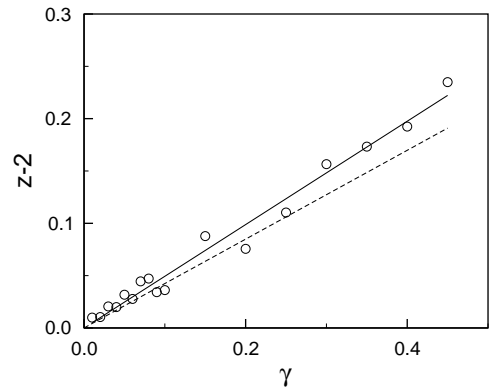}
\caption{Shown are simulation results for the scaling exponent
for the case $n=0$, $\mu = \lambda$, and $\beta = 4$ when torsion is
included.
The error bars are given roughly by the scatter in the data.
The best linear fit to the simulation data is shown (solid line).
 Also shown (dashed line)
is the prediction assuming that the other parameters
do not flow,
$z = 2 + 4 \gamma / (3 \pi)$.
}
\label{fig6}
\end{figure}

For the mathematically interesting case of $n \le 0$ we follow our
previous numerical strategy.  The random displacements of the
diffusing particle are altered from eq.\ (\ref{31}) to
\begin{eqnarray}
\Delta x &=& \frac{D_0 \Delta t }{\sqrt g} \partial_j (\sqrt g g^{1j})
+ D_0 \Delta t g^{j1} T_j
+ w^{1j} z_j
\nonumber \\
\Delta y &=& \frac{D_0 \Delta t }{\sqrt g} \partial_j (\sqrt g g^{2j})
+ D_0 \Delta t g^{j2} T_j
+ w^{2j} z_j \ .
\nonumber \\ 
\label{34}
\end{eqnarray}
To make use of this formula, we need an expression for
$\partial r_i / \partial \sigma_\alpha$ that occurs in $T_j$.
This is found as $\partial r_i / \partial \sigma_\alpha = A^{-1}_{i \alpha}$
where $A_{\alpha i} = \partial \sigma_\alpha / \partial r_i
= \delta_{\alpha i}  - \partial u_\alpha / \partial r_i$.
In evaluating $T_i$, we use the first line of eq.\ (\ref{31aa}).
The results of numerical simulations with this
scheme are shown in Fig.\ \ref{fig6}.
The calculations were performed for the case $\mu = \lambda$, $n=0$, 
$h=1$, and
$\beta  = 4$.  The simulations were done on $4096 \times 4096$ lattices
for a total of 1000000 steps and averaged over 100000 particles.
Also shown is a fit to the functional form of eq.\ (\ref{26a}).
The simulation results are approximately
 fit by $(z- 2)/\gamma =  0.49 \pm 0.04$.

There appears to be relatively little if any renormalization of
$z$ away from the bare value.
A power law behavior of the long-time mean square displacement in the
presence of torsion is observed, although
localization at exceptionally long times still cannot be ruled out.

It is clear that within the continuum assumption of the diffusion
equation, the torsion term affects the dynamics.  The contribution
to the diffusion coefficient is explicit in eq.\ (\ref{33}) for
the case $n>0$.  For $n=0$, the results shown in Fig.\ 
\ref{fig6} differ from those without torsion in Fig.\  \ref{fig5}.
Note that the results with torsion, as those without torsion, differ
substantially from the approximate results of \cite{Turski,Turski2},
noticeably through their dependence on the two Lam{\'e} coefficients.

\section{Conclusion}
We have given a treatment of the effect of topological disorder
on transport properties.  
Within the lattice reconstruction predicted by linear
elasticity theory, topological disorder is
manifestly different from charged, potential-type disorder.
The net effect of the defects, through local lattice expansion and
contraction and global topological rearrangement of lattice connectivity,
is an overall reduction of the transport.
Interestingly, randomly placed dislocations, or randomly placed
disclinations with no net disclinicity,
lead to anomalous subdiffusive behavior when
the displacement fields of linear elasticity are used.

\section*{Acknowledgment}
This research was supported by the Alfred P. Sloan Foundation through
a fellowship to M.W.D.

\bibliography{disloc}

\end{document}